\documentclass[twocolumn]{article}
\usepackage{amsmath,amssymb}
\usepackage[authoryear]{natbib}
\usepackage{graphicx,color,microtype}
\usepackage{subfig}
\usepackage[section]{placeins}
\usepackage[pagebackref]{hyperref}

\begin{document}
\title{Mass entrainment rate of an ideal momentum turbulent round jet}
\author{Ferm\'in Franco Medrano$^{1,2}$\thanks{Corresponding author: franco@kyudai.jp}, Yasuhide Fukumoto$^1$, Clara M. Velte$^3$\\
	and Azur Hod\v zi\'c$^3$\\
	\small{$^1$ Institute of Mathematics for Industry, Kyushu University}\\
	\small{West 1 Bldg., 744 Motooka, Nishi-ku, Fukuoka 819-0395, Japan}\\
	\small{$^2$ Research Institute of Composite Materials, Fukuoka University}\\
	\small{Bldg. 6, 8-19-1 Nanakuma, Jonan-ku, Fukuoka 814-0180, Japan}\\
	\small{$^3$ Department of Mechanical Engineering, Technical University of Denmark}\\
	\small{Bldg. 404, Nils Koppels All\'e, Kgs. Lyngby DK-2800, Denmark}}

\maketitle

\begin{abstract}
	We propose a two-phase-fluid model for a \textcolor{black}{full-cone turbulent round jet} that describes its dynamics in a simple but comprehensive manner with only the apex angle of the cone being a disposable parameter. The basic assumptions are that (i) the jet is statistically stationary and that (ii) it can be approximated by a mixture of \textcolor{black}{two fluids} with their phases in dynamic equilibrium. To derive the model, we impose conservation of the \textcolor{black}{initial} volume and total momentum fluxes. Our model equations admit analytical solutions for the composite density and velocity of the two-phase fluid, both as functions of the distance from the nozzle, from which the dynamic pressure and \textcolor{black}{the mass entrainment rate} are calculated. Assuming a far-field approximation, we theoretically derive a constant entrainment rate coefficient solely in terms of the cone angle. Moreover, we carry out experiments for a single-phase turbulent air jet and show that the predictions of our model compare well with this and other experimental data \textcolor{black}{of atomizing liquid jets}.
\end{abstract}

\section{Introduction}\label{sec:intro}

Liquid jets appear in a vast range of applications. A very active field of application is that of fuel jet injection engines, widely used in the automotive and aerospace industry; other fields of application include medical apparatuses, so-called ``atomizers'' used by commercial products in many industries, flows through hoses and nozzles for various industrial purposes as well as firefighting. Some of these applications and atomization methods are described in \cite{Jiang-etal2010}.

After the pioneering theoretical work of \cite{Rayleigh1878} many other early works advanced and improved upon this subject \citep{Weber1931, Tomotika1935, Taylor1962} \citep[see][]{Gorokhovski2008}. However, these results focused on what is termed ``primary breakup'', i.e. the transition of the jet from a cylindrical geometry to the formation of the first detached droplets and ``ligaments''. This breakup has been found to depend on numerous parameters, and a characterization of the breakup mechanism, based on the dominant physical forces acting on it has been achieved with some success. This was summarized in what is called the ``breakup regimes'' and it can be described broadly in terms of two parameters: the jet's Reynolds and the Weber numbers. Another classification which separated these breakup regimes was based on the jet's speed and the ``Z length'' or the length of the ``continuous'' part of the jet. Four main breakup regimes have been identified: (i) the capillary or Rayleigh regime; (ii) the first wind-induced regime; (iii) the second wind-induced regime; and (iv) the atomization regime. Some review articles on the topic include \cite{McCarthy-Molloy1974, Lin-Reitz1998, Liu2000, Birouk-Lekic2009, Jiang-etal2010}.

The range of applications involving \textit{atomizing} liquid jets forming two-phase fluid flows is still large. The complexity of the atomizing process, involving numerous physical phenomena and many variables, ranging from the conditions inside the nozzle (or some generating source) to the interaction between the atomization process and the environment into which the jet is penetrating, all account for numerous challenges in physical and mathematical modeling. Notwithstanding, several mathematical models have been attempted to describe different aspects of the jets in this regime. For example, differential equations for a fuel jet's tip penetration distance as a function of time \citep{Wakuri1960,Sazhin2001,Desantes-etal05,Pastor-etal08}; models for the gas entrainment rate in a full-cone spray \citep{Cossali}; and a one-dimensional model for the induced air velocity in sprays \citep{Ghosh-Hunt94}. None of these models is sufficient by itself as explained below.

In this study we propose an original 1D mathematical model for the macroscopic dynamics of a full-cone \textcolor{black}{turbulent round jet} ensuing from a circular nozzle into an stagnant \textcolor{black}{fluid}. This kind of jet serves as a basis for many industrial processes in modern manufacturing industry \citep{Jiang-etal2010}. An advantage of our model over other analytical 1D models is that this model has a single experimentally measurable parameter while it maintains reasonable predictive power and gives theoretical understanding that allows it to analytically calculate other physical quantities of interest. Moreover, this model can be extended to accommodate an energy conservation approach with simple turbulence and energy dissipation models, resulting in an increased accuracy as will be reported in a \textcolor{black}{subsequent} paper.

\begin{figure}
	\centering
	\includegraphics[width=\columnwidth]{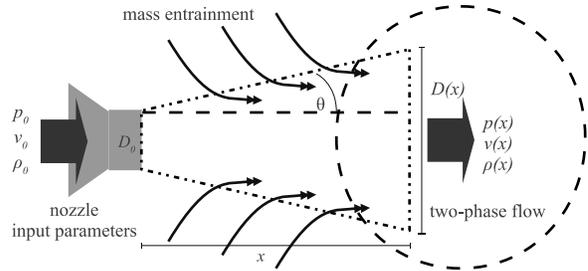}
	\caption{\textcolor{black}{Diagram of the mass entrainment process and the relevant physical variables.}}
	\label{fig:jetdiagram}
\end{figure}

\textcolor{black}{Our particular concern lies in} the \textcolor{black}{\textit{mass entrainment rate}, $\hat m_e$, a dimensionless quantity that describes the proportion of entrained mass in a jet. \textcolor{black}{The mass entrainment process is sketched in Fig. \ref{fig:jetdiagram}.} This quantity is important in applications like diesel engine combustion, as it is related to the rate of mixing of fuel and air \citep{Post-etal2000}. The entrainment rate has been extensively measured for broad experimental settings, as defined from early work on single-phase gas jets \citep{Ricou-Spalding61, Hill72} and later for two-phase liquid-gas jets \citep{Ruff-etal89, Hiroyasu-Arai90, Hosoya-Obokata93, Cossali-etal96}. Formulas derived from data fitting or dimensional analysis are available in the \textcolor{black}{aforementioned} works.
A more recent definition of the \textit{mass entrainment rate} is given by \cite{Post-etal2000} as
\begin{equation}\label{eq:entrainment-rate}
\hat m_e = \frac{\dot m_e(x)}{\dot m_0}=\frac{\dot m(x) - \dot m_0}{\dot m_0}=K_e\left(\frac{x}{D_0}\right)\left(\frac{\rho_e}{\rho_0}\right)^{1/2},
\end{equation}
where $\dot m_e(x)=\dot m(x) - \dot m_0$, and $\dot m_0$ and $\dot m(x)$ are respectively the initial and total \textit{mass flux rates}, defined as the fluid mass fluxes per unit time through a cross-section at respectively axial distance 0 and $x$ from the nozzle. \textcolor{black}{Here} $\rho_0$ and $\rho_e$ are the initial and entrained fluid densities, respectively, $D_0$ is the nozzle diameter and $x$ is the distance from the nozzle's exit.
The dimensionless number $K_e$ is called the \textit{entrainment rate coefficient}, which taking the derivative of Eq. \eqref{eq:entrainment-rate} can thus be calculated as \citep{Post-etal2000}
\begin{equation}\label{eq:Ke}
K_e(x)=\frac{d\dot{m_e}}{dx}\left(\frac{D_0}{\dot{m}_0}\right)\left(\frac{\rho_0}{\rho_e}\right)^{1/2}=\frac{d\hat{m_e}}{d\hat{x}}\rho_*^{-1/2};
\end{equation}
where the last equation uses dimensionless units as defined in Sects. \ref{sec:mathmodel} and \ref{sec:entrainment}. The definition of Eq. \eqref{eq:entrainment-rate} for gaseous jets was motivated by experimental evidence showing that $K_e$ tends to a constant in the far field \citep{Ricou-Spalding61,Hill72,Abraham96,Post-etal2000}, but it can be applied to a non-constant entrainment rate coefficient using Eq. \eqref{eq:Ke}, particularly in the near and intermediate fields.
Despite being originally formulated for gaseous jets, the definition in Eq. \eqref{eq:Ke} is also used for the entrainment rate coefficient of (two-phase) atomized liquid jets \citep{Ruff-etal89,Cossali-etal96,Cossali,Hosoya-Obokata93, Song2003,Rabadi-etal07}. To the the \textcolor{black}{best of the} authors' knowledge however, there is so far no theoretical derivation of it from first principles that is of practical use; although there are some models available \citep{Cossali, Rabadi-etal07}, the most relevant of which are discussed below. One of the goals of this paper is to present this derivation. Our obtained formula generalizes Eq. \eqref{eq:entrainment-rate} and reduces to a constant $K_e$ at large distances from the nozzle (far field). An approximation for the near field is also given.}

In the context of a fuel injection engine, \cite{Sazhin2001} derived a specialized model for the fuel jet's tip penetration distance as a function of time. Their model, however, relied on the unrealistic assumption that the mass ratio of droplets-to-gas ($\alpha_d$, in their notation) is constant throughout the jet's length, which they then used in order to be able to integrate the resulting 1D ODE.
In the same context of a combustion engine and fuel injection, \cite{Desantes-etal05} proposed another 1D model also for the diesel spray tip penetration. Their model relied only on conservation of momentum and on fixing the jet's geometry by means of measuring the cone angle. They replaced the dynamic analysis of a spray by the analysis of an assumed analogously-behaved incompressible gas jet (i.e. ignoring the droplets), by exchanging the original nozzle's exit diameter for an ``equivalent diameter'' and setting the density of the jet equal to that of the gas.
\textcolor{black}{\cite{Pastor-etal08} also proposed a related model for the more general case of the transient (time dependent) evolution of the tip penetration variable. They based their calculations on mass, momentum and enthalpy conservation equations. Their results also present \textcolor{black}{in an appendix for the steady state (time independent) as a special case}. However, even in this special case, their solutions are always numerical, except in a constant-density flow case discussed in the appendix.}
\textcolor{black}{\cite{Cossali}} proposed a model for the gas entrainment in a full cone spray. The model contained newly proposed physical parameters that are not easy to measure experimentally (like three different mean droplet sizes, integrals over the radial distributions of gas and droplets concentrations, velocity, etc.) so it is unclear how the overall theory may produce any prediction from first principles.
Thus, compared to the present study, \textcolor{black}{previous models provide a less comprehensive} description of the density or liquid fraction of the spray, make unrealistic assumptions or introduce parameters unavailable experimentally.

There are also widely used numerical models based on turbulence modeling for jets, like the one by \cite{Vallet2001} used in CFD programs like Star-CD, KIVA-3V, Ansys Fluent, as well as open source codes like OpenFOAM \citep{Stevenin2016}. The present work is concerned with developing an analytical model with the advantage of producing closed expressions.

The article is organized as follows: Sect. \ref{sec:exp} describes the experimental setting. In Sect. \ref{sec:mathmodel} we derive from physical conservation laws the basic mathematical model for the ideal momentum atomizing liquid jet. In Sect. \ref{sec:entrainment} we present the theoretical derivation of the entrainment rate coefficient, based on the derived model, and in Sect. \ref{sec:esp-cases} we apply the model to some common limiting and special cases that recover the known solutions (viz. very thin ambient fluid, (single-phase) submerged jet, and cylindrical jet without breakup). Next, in Sect. \ref{sec:results} we compare the predictions with laboratory experiments from the literature and our own from the Technical University of Denmark. Finally, in Sect. \ref{sec:conclusions} we summarize the main conclusions of the present work.

\section{Experimental method}\label{sec:exp}

The experiments for a single-phase turbulent air jet were carried out at the Department of Mechanical Engineering, Technical University of Denmark. Although this is not an atomizing liquid jet, the special case equations of Sect. \ref{sub:submergedjet} can be used to test the present theory for this case and general insight into how to determine the jet's geometry is obtained (needed for the present model).

\subsection*{Jet facility and enclosure}
The jet was the same as that used by \cite{Ewing2007} and the enclosure was similar as well. The jet was a cubic box of dimensions $58 \times 58.5 \times 59\, cm^3$ fitted with an axisymmetric plexiglas nozzle, tooled into a fifth-order polynomial contraction from an interior diameter of $32\,mm$ to an exit diameter of $D_0 = 10\, mm$. The interior of the box was stacked with foam baffles in order to damp out disturbances from the fan that supplied the generator with pressurized air. For further details on the generator box, see \cite{Jung2004,Gamard2004}.

The flow generating box rested rigidly on an aluminum frame. The exit velocity was monitored via a pressure tap in the nozzle positioned upstream of the contraction and connected to a digital manometer by a silicon tube. The ambient pressure was monitored by an independent barometer. The air intake was located inside the jet enclosure so that both jet and ambient are seeded to provide as homogeneous a seeding distribution in the measured flow as possible. The seeding particles were generated from liquid Glycerin using an atomizing nozzle built in-house, providing tracers with a typical size of about $2-3 \, \mu m$. The enclosure consisted of a large tent of dimension $2.5 \times 3.0 \times 4.5\, m^3$, yielding a cross-sectional area of $2.5 \times 3.0\, m^2$ that resulted in a momentum conservation of $100.0\%$ at $x/D_0 = 30$ and $98.4\%$ at $x/D_0 = 100$, where $x$ is the downstream distance from the jet exit \citep{Hussein1994}.

Depending on ambient conditions, such as barometric pressure and ambient temperature, a pressure drop of about $570\,Pa$ was measured across the jet exit nozzle corresponding to a target jet exit velocity of $30\,m/s$ and Re=20,000.

\subsection*{Measurement equipment}

The ensemble averaged flow field was measured using planar 2-component Particle Image Velocimetry (PIV). The scatterers were illuminated using a DualPower $200-15$ $200\, mJ$ double cavity Nd:YAG $532\, nm$ pulsed PIV laser with maximum power output of $1200\, mJ$ and a pulse duration of $4\, ns$. In order to produce a light sheet following the downstream development of the jet, the laser was positioned below the jet nozzle where an adjustable optical mirror was placed.
\begin{figure}
	\centering
	\includegraphics[width=\columnwidth]{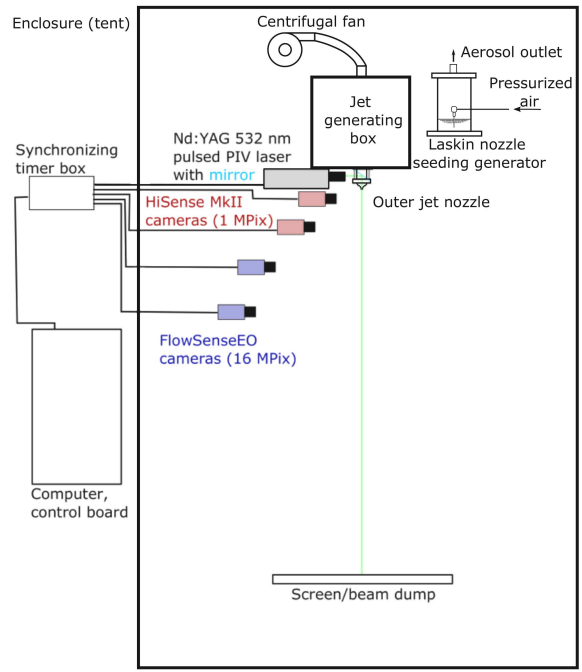}
	\caption{Diagram of the experimental setup and measuring devices.}
	\label{fig:exp_setup_detail}
\end{figure}
Acquisition of the flow field was carried out in four different regions to utilize the spatial resolution capacity of the cameras efficiently. The near field ($x/D_0 = 0.2-6.8$) and the intermediate field ($x/D_0 = 6.8 - 28$) were acquired independently using a HiSense MkII camera with a $1344 \times 1024$ pixels CCD sensor and a $60\, mm$ focal length lens. The aperture was set to $f^{\sharp}2.8$. The far field ($x/D_0= 27.4 -  68.6$ and $66.4 - 106.3$, respectively) was acquired using two Dantec FlowSenseEO $16\, MPix$ cameras ($4872 \times 3248\, Pix$ with a pixel pitch of $7.4\, \mu m$) with $60\, mm$ Nikon lenses with an aperture of $f^{\sharp}2.8$. The different regions sampled and their respective final Interrogation Area (IA) side lengths are listed in Table~\ref{tab:table1}. 
A detailed diagram of the whole setup is shown in Fig. \ref{fig:exp_setup_detail}.
\begin{table}
	\centering
	\caption{Acquisition parameters for the measured regions.}
	\label{tab:table1}
	\begin{tabular}{l|c|c}
		Measured region & extent, $\hat{x}$ & side length, mm\\
		\hline
		Near field & $1.2-6.8$ & 0.80 \\
		Intermediate fld. & $6.8 - 28.0$ & 2.59 \\
		Far field 1 & $27.4 -  68.6$ & 1.69 \\
		Far field 2 & $66.4 - 106.3$ & 2.66
	\end{tabular}
\end{table}

\subsection*{Data processing}

The images were acquired and processed using Dynamic Studio 2015a (4.15.115). A correlation scheme computing $32\times 32$ pixels interrogation areas with $50\%$ overlap in both directions using 2 refinement steps, deforming windows and subpixel interpolation for enhanced accuracy. Local median validation with a $3\times 3$ vector neighborhood validation base was implemented to discard invalid vectors.

Preliminary measurements in the most downstream far field measurement provided an estimate of the integral time scale of $0.059\,s$ from the local jet-half width velocity. In order to ensure statistical independence of the acquired samples, a sampling rate of $1\, Hz$ was chosen. In the near and intermediate fields, $2000$ realizations of the flow were collected to ensure statistical convergence of the mean velocity. Averaging over $500$, $1000$ and $1500$ samples did not display significant deviations from the average from $2000$ realizations and the statistics can therefore be considered converged to sufficient degree for the current purposes. For the far field, 11,000 realizations were collected.

In relevant cases, to remove faint reflections and other adverse effects of the background, the images were first preprocessed by computing, for each pixel separately, the minimum intensity of the ensemble of measurement images and subtracting this from the same ensemble. The average velocity field was then obtained by only including valid vectors, discarding vectors substituted by the validation algorithm of the software.

\section{The ideal momentum atomizing liquid jet model}\label{sec:mathmodel}

Consider a stationary full-cone \textcolor{black}{turbulent round jet} ensuing from a circular nozzle (of small diameter) into an ambient \textcolor{black}{fluid}, with a constant high gauge pressure \textcolor{black}{input fluid} and a small conical jet angle. We want to calculate the dynamical variables of the two-phase jet at some distance from the nozzle. The relevant variables and parameters of the model are depicted in Fig. \ref{fig:jetdiagram}, where $\theta$ is the spread half-angle of the conical jet, $\rho$ the density of the fluid, $D$ the diameter of jet, \textcolor{black}{$p$ the fluid's pressure}, $v$ the axial velocity of a fluid element averaged over a cross section of constant $x$, where $x$ is the axial distance from the nozzle. The subscript ``0'' indicates initial values (at the nozzle's exit position), e.g. $D_0$ is the diameter of the jet at the nozzle's exit (equal to the nozzle's orifice diameter).

\subsection{Initial momentum flux}\label{sub:initial-momentum}

The momentum of a flat disc of \textcolor{black}{fluid} of infinitesimal width exiting the circular nozzle is
\begin{equation}\label{eq:p0}
d\Pi_0=m_0v_0,
\end{equation}
where the mass of the flat disc is $m_0=\frac{1}{4}\rho_0\pi D_0^2\,dx$. Note that $dx=v_0\,dt$. This velocity may be calculated from the input gauge pressure, $p_0$, by Bernoulli's theorem, neglecting the dynamic pressure inside the nozzle and assuming the static pressure is totally converted to the jet's dynamic pressure just outside the nozzle. Thus $p_0\approx\frac{1}{2}\rho_0v_0^2$ implies
\begin{equation}\label{eq:v0}
v_0=\sqrt{\frac{2p_0}{\rho_0}}.
\end{equation}
Substituting $m_0$ and $v_0$ into equation (\ref{eq:p0}) we get
\begin{equation}\label{eq:p0flux}
\dot{\Pi}_0:=\left.\frac{d\Pi}{dt}\right|_{x=0}=\frac{1}{2}\pi D_0^2 p_0,
\end{equation}
which is the momentum flux per unit time coming out of the nozzle as a result of the input gauge pressure inside the nozzle.

\subsection{Conservation of momentum}

\textcolor{black}{We assume that the fluid at a distance $x$ is a two-phase fluid.} The present model has been termed ``ideal momentum'' to distinguish it from ``ideal energy'' and \textcolor{black}{``lossy''} models including momentum, energy or mass loss parameters, developed by the present authors and to be reported in detail in a subsequent article. Accordingly, first assume conservation of momentum, i.e. the momentum of the two-phase fluid is solely that from the original input pressure. The droplets transfer momentum to the initially static \textcolor{black}{fluid} by drag forces \citep{Sazhin2006, Fuchimoto-etal09} and they reach local dynamic equilibrium  in such a way that both \textcolor{black}{phases} move at the same speed $v$ inside the jet \citep{Desantes-etal11}; this is also the main assumption under the wide class of ``Locally Homogeneous Flows'' (LHF) \citep{Faeth83, Faeth87, Faeth-etal95}. We assume that the latter process occurs so fast immediately outside the nozzle's exit that we may neglect the non-equilibrium zone near the nozzle. The latter assumption is reasonable for high-speed pressure atomized jets, e.g. like the ones used in real life diesel engines \citep{Sazhin2001}. \textcolor{black}{This effectively allows us to treat the two-phase flow as a \textit{single fluid with a composite density $\rho(x)$ and a single velocity $v(x)$}, depending on the distance from the nozzle, $x$.}

Analogous to the calculations in Sect. \ref{sub:initial-momentum}, the momentum at some distance $x$ is $d\Pi=mv$, where $m=\frac{1}{4}\rho\pi D^2\,dx$ and $dx=v\,dt$, by which
\textcolor{black}{\begin{equation}\label{eq:p}
\frac{d\Pi(x)}{dt}=\frac{1}{4}\pi\rho D^2 v^2,
\end{equation}}
which is the momentum flux at a cross-section of the two-phase fluid jet at a distance $x$ from the nozzle.

Equating Eq. \eqref{eq:p0flux} and Eq. \eqref{eq:p} by conservation of the momentum contained between the two planes $x=0$ and at axial distance $x$ in a steady state, using again Eq. \eqref{eq:v0} and solving for $v$ we get
\begin{equation}\label{eq:vi}
v^2=\left(\frac{D_0}{D}\right)^2\frac{\rho_0}{\rho}v_0^2.
\end{equation}
This can be written in simpler form by using dimensionless units as $\hat{v}^2=\hat{\rho}^{-1}\hat{D}^{-2}$, where $v$, $D$ and $\rho$ have been scaled respectively by $v_0$, $D_0$ and $\rho_0$ (the quantities at the nozzle's exit) and denoted by a hat symbol.
\subsection{\textcolor{black}{Volume of entrained mass in the two-phase fluid}}

The total fluid volume of the thin disc at the target distance is \textcolor{black}{$dV=dV_0+dV_e$, where the subscript ``e'' denotes the quantities related to the entrained fluid}, i.e. the total volume $dV$ of the two-phase fluid is just the sum of the volume of the original quantity of fluid coming out of the nozzle $dV_0$ plus the added volume of entrained \textcolor{black}{fluid} in dynamic equilibrium, \textcolor{black}{$dV_e$.} Also, the total volume of the \textcolor{black}{jet} at the target distance is straightforward to calculate from the conical geometry. Then \textcolor{black}{$dV_e=dV-dV_0$} from where we can calculate a volumetric flow rate of the \textcolor{black}{mass} entrainment:
\textcolor{black}{\begin{equation}\label{eq:entrainment}
\frac{dV_e}{dt}=\frac{1}{4}\pi(D^2v-D_0^2v_0) 
\end{equation}}
Here we can substitute for $v_0$ from equation (\ref{eq:v0}) and $D=D_0+2x\tan(\theta)$.

\subsection{Density of the two-phase fluid}

The mean composite density of a two-phase fluid thin-disc element is just the total mass over the total volume:
\textcolor{black}{\begin{equation}\label{eq:meandens}
\rho=\frac{dm}{dV}=\frac{dm_0+dm_e}{dV_0+dV_e}=\frac{\rho_0dV_0+\rho_edV_e}{dV_0+dV_e}.
\end{equation}}
After substituting \textcolor{black}{$dV_e$} from Eq. \eqref{eq:entrainment}, and the initial volume element $dV_0$ from analogous calculations, simplifying and solving for $\rho$, we obtain
\textcolor{black}{\begin{equation}\label{eq:ri}
\rho=\rho_e+\frac{D_0^2v_0}{D^2v}(\rho_0-\rho_e)\quad\textrm{or}\quad\hat{\rho}=\rho_*+(1-\rho_*)/\hat{D}^2\hat{v},
\end{equation}}
the density of the two-phase fluid at the distance $x$ from the nozzle. \textcolor{black}{Fortunately, it depends on $v$ which makes the dependency closed as we can see from Eq. \eqref{eq:vi}.} The second equation in Eq. \eqref{eq:ri} is the dimensionless form, where \textcolor{black}{$\rho_*=\rho_e/\rho_0$}.

We implicitly assume, by calculating the \emph{mean composite density} of the two-phase fluid in Eq. \eqref{eq:meandens}, that the droplets distribution throughout the disc two-phase fluid element does not differ greatly from a uniform distribution. Notice that we approximate the front of the jet by a planar front of equal density, i.e. a ``top-hat'' radial distribution. In reality this is not true, since the front should be spherical in the first order, and it is then in a spherical shell within the jet's cone that we should consider $\rho$ to be approximately constant, not in a plane. However, for small half-angles $\theta$ and short distances $x$ a plane should be suffice as a first order approximation. The same could be said of the front's velocity $v$. Overall, we may take the above considerations as utilizing ``top-hat'' velocity and density distributions as a first approximation. Note that slicing the spherical jet front with a $x$-normal plane provides a non-constant $\rho$ density distribution in this plane. This distribution should, however, be similar to a two-dimensional Gaussian distribution centered around the $x$-axis, i.e. the jet's ``centerline''. There are some models \citep{Ghosh-Hunt94,Desantes-etal05, Rabadi-etal07, Pastor-etal08,Desantes-etal11} which apply \textcolor{black}{self-similar} Gaussian velocity distributions as initial assumptions; however, this calculation will be included in a later work since we anticipate that it would not lead to a major refinement of the axial centerline quantities.

\subsection{Explicit expressions for the axial velocity and composite density of the spray}

From Eqs. \eqref{eq:vi} and \eqref{eq:ri} we identify a system of two nonlinear equations with two unknowns, $v$ and $\rho$. We can eliminate $\rho$ from the system to get $v$ explicitly in dimensionless form as
\begin{equation}\label{eq:vhat}
\hat{v}=\frac{\rho_*-1+\sqrt{(\rho_*-1)^2+4\hat{D}^2\rho_*}}{2\hat{D}^2\rho_*}.
\end{equation}
\textcolor{black}{Eq. \eqref{eq:vhat} is equivalent to the one derived independently by \cite{Wakuri1960} using a similar approach.} Analogously, we obtain an explicit dimensionless form for the density, $\hat{\rho}=\rho/\rho_0$, by eliminating $v$ from the same described system, obtaining:
\begin{equation}\label{eq:rhat}
\hat{\rho}=\tilde{\rho}+\sqrt{\tilde{\rho}^2-\rho_*^2},
\end{equation}
where $\tilde{\rho}=\rho_*+(1-\rho_*)^2/2\hat{D}^2$. From Eqs. \eqref{eq:vhat} and \eqref{eq:rhat}, the dimensionless form of the dynamic pressure, \textcolor{black}{$\hat{p}=p/p_0=\hat{\rho}\hat{v}^2$} (where $p_0=\frac{1}{2}\rho_0v_0^2$ is the output dynamic pressure at the nozzle's exit), which accounts for the total pressure at the target distance, may be calculated as
\begin{equation}\label{eq:phat}
\hat{p}=\left(\frac{\hat{\rho}}{\hat{D}}\right)^2.
\end{equation}
Notice the implicit single-variable dependence of $\hat{v}(\hat{x})$, $\hat{\rho}(\hat{x})$ and $\hat{p}(\hat{x})$ on the \textcolor{black}{dimensionless axial distance $\hat{x}=x/D_0$} by substituting $\hat{D}=1+2\hat{x}\tan\theta$ in Eqs. \eqref{eq:vhat}-\eqref{eq:phat}.

\section{\textcolor{black}{Mass entrainment rate}}
\label{sec:entrainment}

\textcolor{black}{We have the \textit{volumetric} entrainment rate \textcolor{black}{$dV_e/dt$} as given by Eq. \eqref{eq:entrainment}; which multiplying by $\rho_e$ and normalizing by the volumetric flux at the nozzle, $\dot{m}_0= \frac{1}{4}\pi\rho_0 D_0^2 v_0$, we get the \textcolor{black}{mass entrainment rate}. Using the dimensionless notation introduced above, we can write}
\begin{equation}\label{eq:me}
\hat{m}_e=\rho_*(\hat{D}^2\hat{v}-1).
\end{equation}
\textcolor{black}{as defined in Eq. \eqref{eq:entrainment-rate}. Notice $\frac{d\hat{V}_g}{dt}=\hat{D}^2\hat{v}-1$, by which \eqref{eq:me} can be written simply as $\hat{m}_e=\rho_*\frac{d\hat{V}_e}{dt}$.}
Substituting $\hat{v}$ from Eq. \eqref{eq:vhat} we have
\begin{equation}\label{eq:me-explicit}
\hat{m_e}(\hat{x})=\frac{1}{2}\left[\sqrt{(\rho_*-1)^2+4\rho_*\hat{D}(\hat{x})}-\rho_*-1\right].
\end{equation}
Equation \eqref{eq:me-explicit} gives us an explicit relationship between the mass entrainment rate $\hat{m_e}$ and the axial distance, $x$, remembering that $\hat{D}(\hat{x})$ depends solely \textcolor{black}{on} this distance and the jet's spread angle, $\theta$. \textcolor{black}{It is thus a theoretical derivation of Eq. \eqref{eq:entrainment-rate} in a generalized form and from first principles.}
It is easy to check that when $x\rightarrow0$, i.e. at the nozzle's exit, $\hat{m}_e\rightarrow0$, as it should be since there is no entrained gas there yet.

Taking the derivative of Eq. \eqref{eq:me} with respect to $\hat{x}$ and substituting $\hat{v}$ from \eqref{eq:vhat}, we get
\begin{equation}\label{eq:dmedx}
\frac{d\hat{m_e}}{d\hat{x}}=\frac{4\rho_*\hat{D}\tan\theta}{\sqrt{(\rho_*-1)^2+4\rho_*\hat{D}^2}}.
\end{equation}
We can transform Eq. \eqref{eq:dmedx} into its dimensional form and \textcolor{black}{substitute it into Eq. \eqref{eq:Ke}} to obtain a new explicit formula for $K_e$, which can then be written as
\begin{equation}\label{eq:Ke-general}
K_e(x)=\frac{4\rho_*^{1/2}D(x)\tan\theta}{\sqrt{(\rho_*-1)^2 D_0^2+4\rho_*D(x)^2}}.
\end{equation}
Equation \eqref{eq:Ke-general} is a \textcolor{black}{new} general explicit expression for the \textit{mass} entrainment rate coefficient as a function of the axial distance from the nozzle. Notably, $K_e(x)$ includes the half angle $\theta$ of the cone apex as an adjustable parameter. In Sect. \ref{sub:near-far} below we will theoretically \textcolor{black}{deduce} that our formula in Eq. \eqref{eq:Ke-general} can be approximated by a constant \textcolor{black}{for the far field, as proposed by other authors \citep{Ricou-Spalding61,Hill72,Abraham96,Post-etal2000} based on experimental results. We also will give a new constant value for the near field.}

\section{Limiting and special cases}\label{sec:esp-cases}

Let us now calculate some special and limit cases of the ideal momentum atomizing liquid jet model.

\subsection{The near and far fields}
\label{sub:near-far}

Very near the nozzle, taking $\hat{x}\rightarrow0$ in Eq. \eqref{eq:vhat} we get $\hat{D}\rightarrow 1$ and $\hat{v}\rightarrow1$, correspondingly $v\rightarrow v_0$ in dimensional form, as expected since $v_0$ is the velocity of the liquid exiting the nozzle. From Eq. \eqref{eq:rhat} we get $\hat{\rho}\rightarrow1$ thus $\rho\rightarrow\rho_0$, also as expected. We can also check from Eq. \eqref{eq:phat} that $\hat{p}\rightarrow 1$, thus $p\rightarrow p_0$. All the limiting values agree with the ones inside the nozzle. As for the entrainment rate, taking the limit of Eq. \eqref{eq:Ke-general} for $x\rightarrow0$ we have the \textcolor{black}{\textit{near field entrainment rate constant}}
\begin{equation}\label{eq:Ke-near}
K_{e,\,near}=\frac{4\rho_*^{1/2}\tan\theta}{\rho_*+1},
\end{equation}
because $D(x)\rightarrow D_0$ as $x\rightarrow0$. On the other hand, in the far limit, taking $\hat{x}\rightarrow+\infty$ in \eqref{eq:vhat} we get $\hat{v}\rightarrow0$ which also corresponds to $v\rightarrow 0$ in dimensional form, as expected since the ideal jet expands infinitely, $\hat{D}\rightarrow+\infty$, and thus eventually tends to diffuse all its kinetic energy to the increasing amounts of entrained gas. From Eq. \eqref{eq:rhat} we get $\hat{\rho}\rightarrow\rho_*$ and thus $\rho\rightarrow\rho_e$, also as expected, since the initial amount of \textcolor{black}{fluid} should diffuse with the ever-increasing amounts of entrained gas and eventually the composite density tend to the one of the latter. From Eq. \eqref{eq:phat} we get $\hat{p}\rightarrow 0$, as does $p$ then, which follows from the zero velocity. For the entrainment rate coefficient, taking the \textcolor{black}{approximation} of Eq. \eqref{eq:Ke-general} for $x\gg D_0$, i.e. $\hat{D}\gg1$, we get the \textcolor{black}{\textit{far field entrainment rate constant}}
\begin{equation}\label{eq:Ke-far}
K_{e,\,far}=2\tan\theta.
\end{equation}
Considering a half-angle of $8^\circ\lesssim \theta\lesssim10^\circ$ as found in the experiments by \cite{Hiroyasu-Arai90} for high-speed (over 70 m/s) pressure-atomized liquid jets, we get $0.28\lesssim K_{e,far}\lesssim0.35$. Remarkably, the value 0.32 used by \cite{Ricou-Spalding61} for gaseous jets in their experiments falls into the range above, suggesting a half-angle of $9^\circ$.

By comparing Eqs. \eqref{eq:Ke-near} and \eqref{eq:Ke-far}, we have
\begin{equation}\label{eq:Ke-rel}
K_{e,near}=\frac{2\rho_*^{1/2}}{\rho_*+1}K_{e,\,far}.
\end{equation}
We can check that the coefficient of $K_{e,far}$ in the right-hand-side of Eq. \eqref{eq:Ke-rel} is always positive and less than one since \textcolor{black}{$\rho_*>0$}. Therefore we prove that $K_{e,near}<K_{e,far}$, implying that the entrainment rate has a tendency of growing with the distance from the nozzle, independently of the density ratio $\rho_*$. \textcolor{black}{We have thus confirmed that Eq. \eqref{eq:Ke-general} is a generalization of the empirical one given by Eq. \eqref{eq:entrainment-rate},} and provides a theoretical account for it.

\subsection{Liquid jet without breakup}

In this case, the liquid jet does not disintegrate, i.e. it does not atomize, and the jet consists of pure liquid. This corresponds to a spread half-angle zero, $\theta=0$. Taking $\theta\rightarrow0$ in Eq. \eqref{eq:vhat} and thus $\hat{D}\rightarrow1$, we immediately see that $\hat{v}\rightarrow 1$, so that $v\rightarrow v_0$, which has the physical interpretation that the liquid keeps propagating at the same initial speed. As for the density, we see from Eq. \eqref{eq:rhat} that $\rho\rightarrow \rho_0$ since $\hat{\rho}\rightarrow1$, which again agrees with the physical meaning as we should only have liquid. $\hat{p}\rightarrow1$ so that the dynamic pressure is unaltered along the jet's flight. Finally it is immediate to see from Eq. \eqref{eq:Ke-general} that if $\theta=0$ then $K_e=0$ as expected.

\subsection{The ideal liquid jet in a very thin atmosphere}

In its dimensionless form, this condition means $\rho_*\approx 0$. With this assumption, $\rho_e\approx0$ or $\rho_e\ll\rho_0$, which physically means a liquid jet inside a very thin ambient gas. Taking the limit in Eq. \eqref{eq:vhat} we see that $\hat{v}\rightarrow 1$, which corresponds to $v\rightarrow v_0$; i.e. the velocity remains constant at the same initial value, which is physically acceptable in the sense that liquid droplets would not loose any momentum to the surrounding thin ambient gas since there would be little drag forces present.
As for the density, taking the same limit in Eq. \eqref{eq:rhat}, we see that $\tilde{\rho}\rightarrow1/2\hat{D}^2$ implies $\hat{\rho}=\hat{D}^{-2}$, and $\hat{p}=\hat{D}^{-6}$. The entrainment rate in this case is also zero as can be verified from \eqref{eq:Ke-general} by taking the limit $\rho_*\rightarrow0$; this is physically consistent as well.

From the above, in dimensional form, the \textit{state equations for an ideal momentum atomizing liquid jet in a thin atmosphere} are then
\begin{eqnarray}\label{eq:r-vacuum}
v=v_0, \quad \rho=\rho_0\left(\frac{D_0}{D_0+2x\tan\theta}\right)^2,\\
p=p_0\left(\frac{D_0}{D_0+2x\tan\theta}\right)^6, \quad \textrm{and } K_e(x)=0;
\end{eqnarray}
where we can see that as $x\rightarrow 0$, $\rho\rightarrow\rho_0$ and $p\rightarrow p_0$, accordingly. As $x\rightarrow\infty$, $\rho\rightarrow 0$ and $p\rightarrow0$, the approximately null density and dynamic pressure of the thin gas, respectively.

\subsection{The submerged jet}\label{sub:submergedjet}

This is a single-phase jet, e.g. liquid-liquid or gas-gas jet; correspondingly, we should have $\rho_e=\rho_0$. This case corresponds to $\rho_*=1$ in dimensionless form. From Eq. \eqref{eq:rhat} we have that $\rho=\rho_0(\equiv\rho_e)$, as expected so the density is just that of the common fluid for all distances. \citeauthor{Landau} called this a ``submerged jet'' while \citeauthor{Batchelor} called it a ``point source of momentum jet''.

From Eqs. \eqref{eq:vhat}, \eqref{eq:phat} and \eqref{eq:Ke-general} we get that for this case $\hat{v}=\hat{D}^{-1}$, $\hat{p}=\hat{D}^{-2}$ and $K_e(x)=2\tan\theta$; so that, in dimensional form, the \textit{state equations for the ideal momentum submerged jet} are
\begin{eqnarray}\label{eq:mom-jet}
v=v_0\left(\frac{D_0}{D_0+2x\tan\theta}\right), \quad\rho=\rho_0,\\
p=p_0\left(\frac{D_0}{D_0+2x\tan\theta}\right)^2,\quad\textrm{and } K_e=2\tan\theta;
\end{eqnarray}
which for $x\rightarrow0$, i.e. at the nozzle's exit, still gives $v=v_0$ and $p=p_0$, while for $x\rightarrow\infty$, $v=p=0$, \textcolor{black}{as is physically consistent}. Notice that in this case density remains constant and velocity decays with distance whilst for the case of a thin ambient gas we had the converse, i.e. the velocity remains constant and the density decays. Also, the entrainment rate coefficient is constant and $K_e=K_{e,far}$ for all distances, which is the same as the constant coefficient defined by \cite{Ricou-Spalding61}. This can be interpreted from Eq. \eqref{eq:Ke-rel} where we can see that, while on the near-field the effect of the density ratio is considerable, in the far field it is not present. This is because the mass flux through a cross-section close to the nozzle is highly influenced by the density of the fluid exiting it, while this effect gradually decreases becoming negligible in the far-field, as the mass flux of the original fluid is small compared to the entrained mass flux.

Finally, if we substitute $v_0$ from Eq. \eqref{eq:v0} into Eq. \eqref{eq:mom-jet} we obtain
\textcolor{black}{\begin{equation}\label{eq:Landau1}
	v=2\pi^{-1/2}\left(\frac{\dot{\Pi}_0}{\rho}\right)^{1/2}(D_0+2x\tan\theta)^{-1}
\end{equation}}
where $\dot{\Pi_0}=d\Pi_0/dt$ is the momentum flux at the nozzle exit given in Eq. \eqref{eq:p0flux}. For $x\gg D_0$ we can drop the $D_0$ in the last factor of Eq. \eqref{eq:Landau1} obtaining the following proportionality relationship:
\begin{equation}\label{eq:Landau}
	v\propto x^{-1}\sqrt{\dot{\Pi_0}/\rho},
\end{equation}
which is the classical result for the mean velocity decay of a turbulent jet due to \cite{Landau}.

\section{Comparison with experiments and discussion}\label{sec:results}

We will show the agreement of the ideal atomizing liquid jet model with two sets of experimental results for the entrainment rate coefficient of two-phase liquid-gas jets and two more sets for the axial velocity of a single-phase air jet.

\subsection{Atomizing diesel jet in air}
\begin{figure}
	\centering
	\includegraphics[width=\columnwidth]{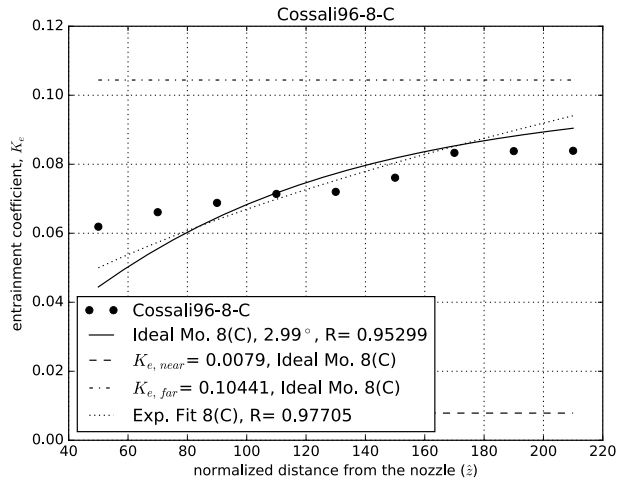}
	\caption{\textcolor{black}{Air entrainment coefficient of a steady pressure-atomized diesel fuel jet entering stagnant ambient air. Experimental results by \cite{Cossali-etal96}, far field only ($\hat{x}>30$). Pressure drop was 21.4 MPa and ambient air temperature was 298$^\circ K$. The optimized jet's half-angle parameter is $\theta=2.99\,^\circ$ with a correlation coefficient of $R\approxeq0.953$.}}
	\label{fig:cossali96}
\end{figure}

\cite{Cossali-etal96, Cossali} reported experimental results for the air entrainment coefficient of two-phase pressure-atomized diesel jets in air inside pressure and temperature-controlled chambers, approximating the conditions inside a fuel-injection combustion engine. For all experiments they used single-hole small opening nozzles with $D_0=0.25$ mm, using standard diesel fuel of density $\rho_0=820\,\mathrm{kg/m^3}$. Of these, only one experiment was performed with a chamber at an ambient pressure of 101.325 kPa (1 atm) and an chamber temperature equal to an ambient temperature of 25$^\circ$C, denoted as experiment ``C''. At this temperature, $\rho_e\approx1.169\,\mathrm{kg/m^3}$. The injection pressure is reported therein to have been 21.4 MPa (214 bar). Using Eq. \eqref{eq:v0} we can estimate $v_0\approx228.5$ m/s and consequently $\dot{m}_0\approx9.1$ g/s. 
The results are shown in Fig. \ref{fig:cossali96}. The solid line is Eq. \eqref{eq:Ke-general}, using a parameter $\theta=2.99^\circ$ optimized by least squares fitting of the curve. In this case, the trend agrees with the experimental data for the far field, viz. \textcolor{black}{$\hat{x}\gtrsim30$ with a correlation coefficient of $R = 0.95299$}. It must be noted that the best experimental curve fit available for the same data, as given by the original authors themselves \citep{Cossali-etal96}, is not much closer (this curve is not shown in their article), \textcolor{black}{having a correlation coefficient of $R=0.97705$}. The dotted curve in Fig. \ref{fig:cossali96} labeled ``Exp. Fit 8(C)'' corresponds to this experimental power-law fit given in \cite{Cossali-etal96} as
\begin{equation}\label{eq:expKe}
	K_e(x)=B\left(\frac{x+x_0}{D_0}\right)^{0.5} \left(\frac{\rho_e}{\rho_0}\right)^{0.31} \left(\frac{T_g}{T_0}\right),
\end{equation} 
where $B=0.044$ is some scaling parameter, $x_0$ is the position of the nozzle's exit (defined when $x=0$) and $T_g$ and $T_0$ are the temperatures of the gas and liquid, respectively. The exponents of the power law were calculated using an optimized fit of the parameters in a RMS sense, adjusting not only for this but for the data of several experiments altogether. This experimental power-law seems to approximate better in other cases presented by the same authors.

Finally, we must note that the obtained optimized angle $\theta\approx 3^\circ$ is smaller than the angle expected for the whole jet, estimated between about $8^\circ\lesssim \theta\lesssim10^\circ$ as in the experiments by \cite{Hiroyasu-Arai90}. We must, however, note that there are great variations in the angle depending on the specific jet characteristics, most notably the densities ratio $\rho_*$ \citep{Lefebvre89,Hiroyasu-Arai90,Desantes-etal05,Xie2015,Emberson2016}. The constant-value approximations for the entrainment rate coefficient in the near and far fields are also shown in Fig. \ref{fig:cossali96} using the same angle in Eqs. \eqref{eq:Ke-near} and \eqref{eq:Ke-far}. Here we see that $K_{e,\,far}$ agrees relatively well with the far field data, while $K_{e,\,near}$ more notably underestimates the experimental value in this case.
 
\subsection{Atomizing water jet in air}
\begin{figure}
	\centering
	\includegraphics[width=\columnwidth]{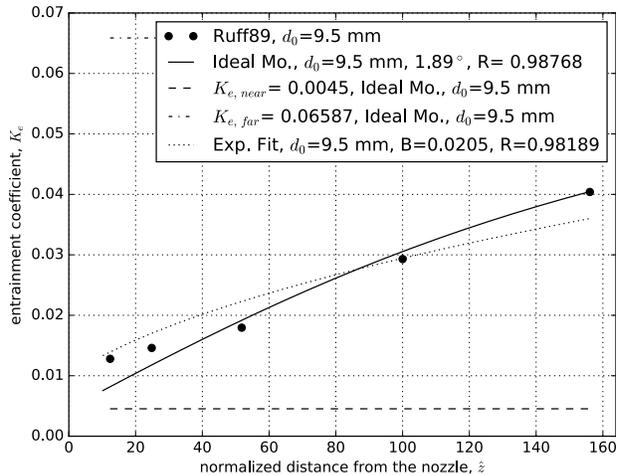}
	\caption{\textcolor{black}{Air entrainment coefficient of a steady pressure-atomized water jet entering stagnant ambient air. Experimental data from \cite{Ruff-etal89}, intermediate and far fields only ($\hat{x}>10$). $D_0=9.5$ mm and pressure drop of 2.52 MPa. The parameter $\theta=1.89$ is the optimized value, obtaining a correlation coefficient of $R\approxeq0.988$.}}
	\label{fig:ruff89}
\end{figure}

\cite{Ruff-etal89} reported experimental results for a pressure-atomized water jet entering stagnant ambient air, injected downward using a nozzle with $D_0=9.5$ mm, a fuel density of $\rho_0$ = 998 kg/m$^3$, ambient temperature $T_g=298\pm 2 ^{\circ}$K, reporting for water a Reynolds number (Re = $\rho vD/\mu$, where $\mu$ is the kinematic viscosity) Re = 5.34$\times10^{5}$ and Weber number (We = $\rho v^2D/\sigma$, where $\sigma$ is the surface tension \textcolor{black}{coefficient}) of We = 411.5$\times 10^3$, while for air We = 492.8. 
The results for the air entrainment coefficient using a pressure drop of 2.52 MPa are shown in Fig. \ref{fig:ruff89} along with our model fit, Eqs. \eqref{eq:Ke-general}, \eqref{eq:Ke-near} and \eqref{eq:Ke-far}, using a jet's half angle parameter of $\theta=1.89^{\circ}$ \textcolor{black}{and obtaining a correlation coefficient $R=0.98768$}. We estimate $v_0=71.05$ m/s from Eq. \eqref{eq:v0}. The power-law experimental fit by \cite{Cossali-etal96} is also shown in the same figure \textcolor{black}{getting a correlation coefficient $R=0.98189$, slightly lower than with our model}. It must be noted, however, that in their paper Cossali et al. gave $B=0.044$, but using this value gives a very bad fit for the presently considered experiment by \cite{Ruff-etal89}. Therefore, we recalculated the value of this parameter by least-squares optimization, obtaining $B=0.0204$. Notably our model seems to adjust very well in the far field, while still not being so far in the \textcolor{black}{intermediate} field. The experimental power-law follows an opposite trend in this case, giving a better fit in the near field but differing in the far field. $K_{e, near}$ underestimates the experimental values in the near field but is still a reasonable approximation. $K_{e, far}$ does seem to overestimate the far field trend considerably. We must remark that the optimized value of the half-angle, $\theta\approx2^\circ$, is again smaller than the expected value for the whole jet geometry \citep{Hosoya-Obokata93}. However, which angle to use as a parameter is not clear, as there is no single universally agreed definition of $\theta$ \citep{Lefebvre89,Hiroyasu-Arai90,Desantes-etal05,Xie2015,Emberson2016}. We will give our own original approach to this issue in Sect. \ref{sub:airjet} below.

\subsection{Single-phase air jet}\label{sub:airjet}

\begin{figure*}[t]
	\centering
	\includegraphics[width=0.325\textwidth]{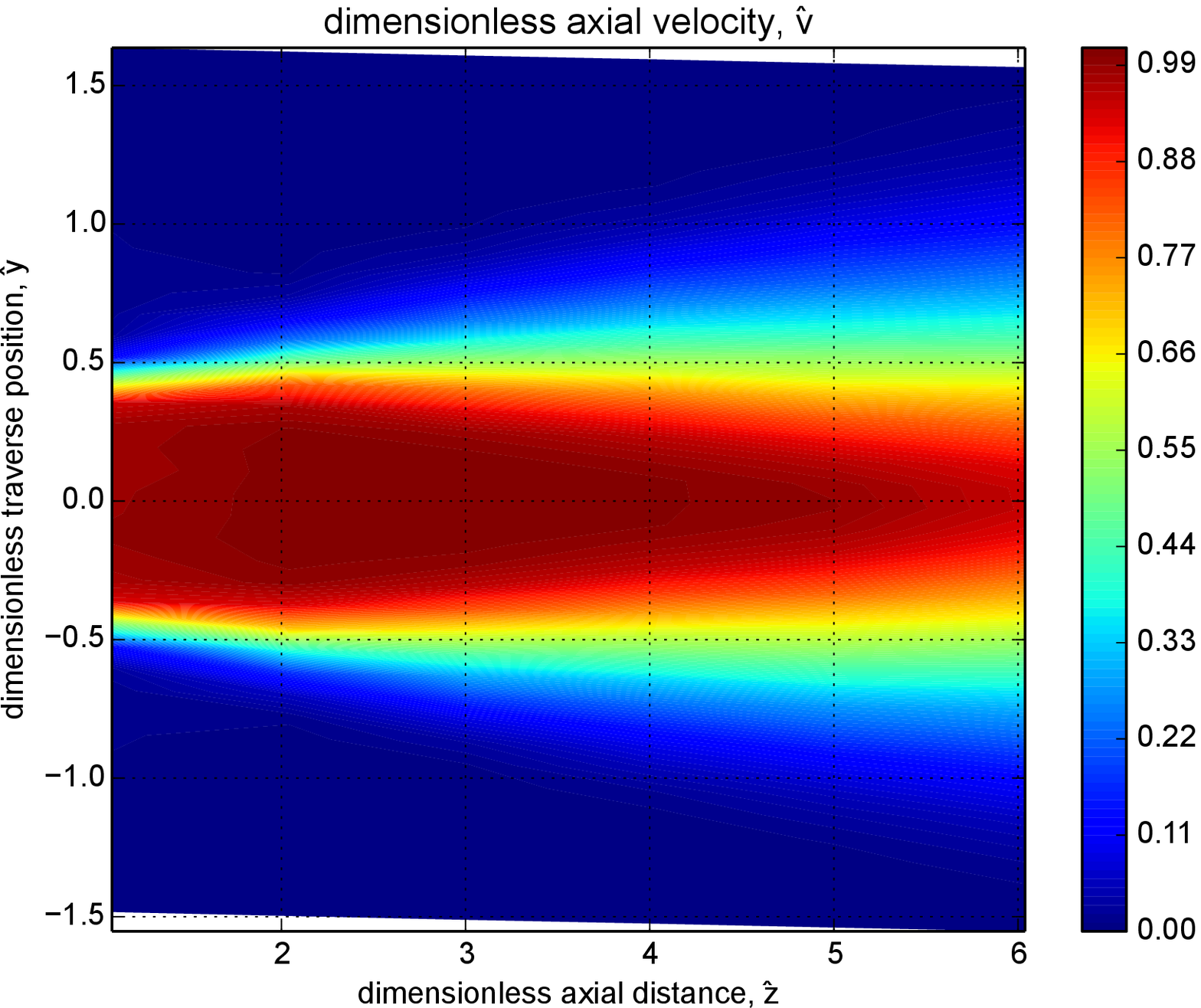}
	\includegraphics[width=0.325\textwidth]{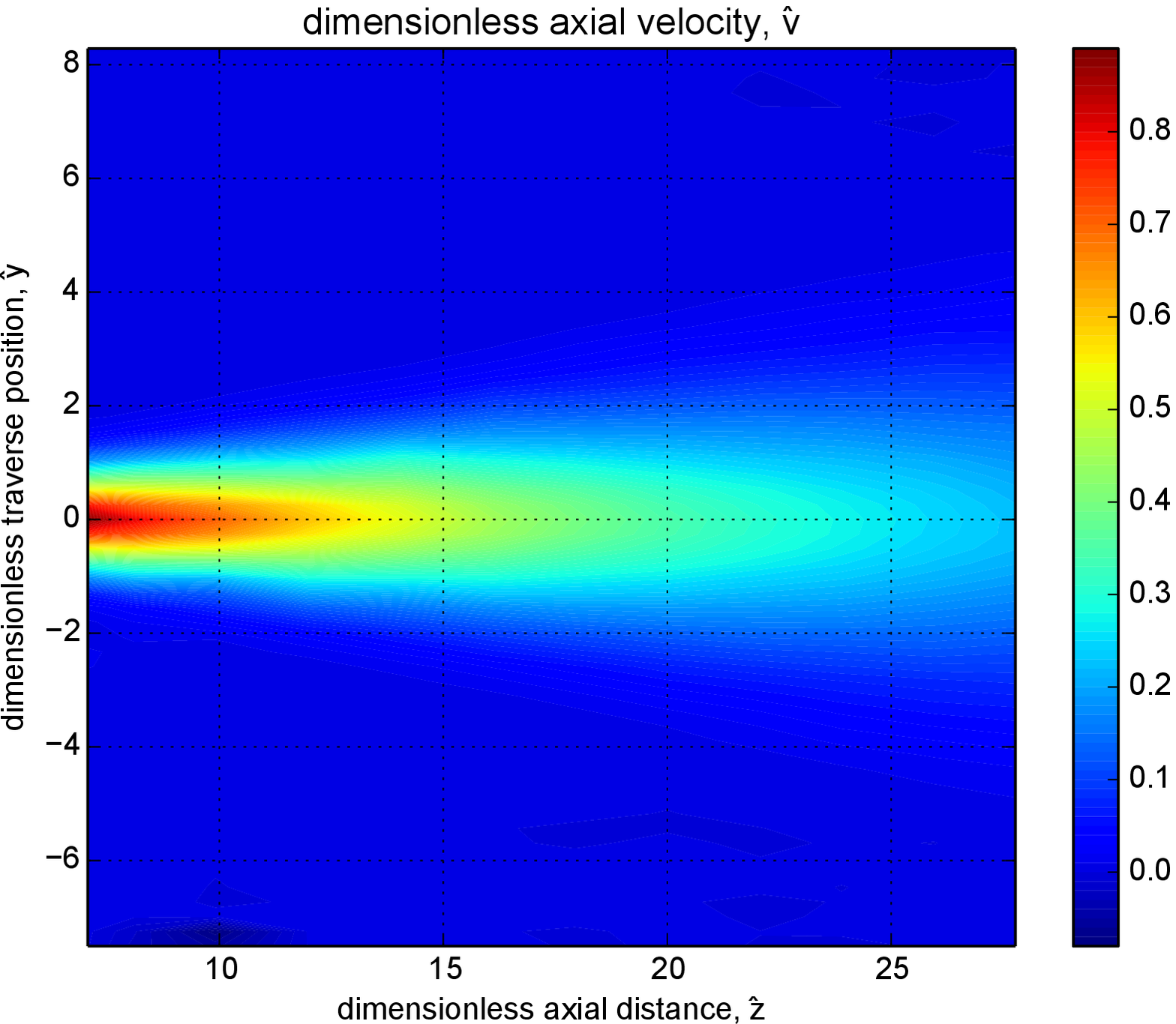}
	\includegraphics[width=0.325\textwidth]{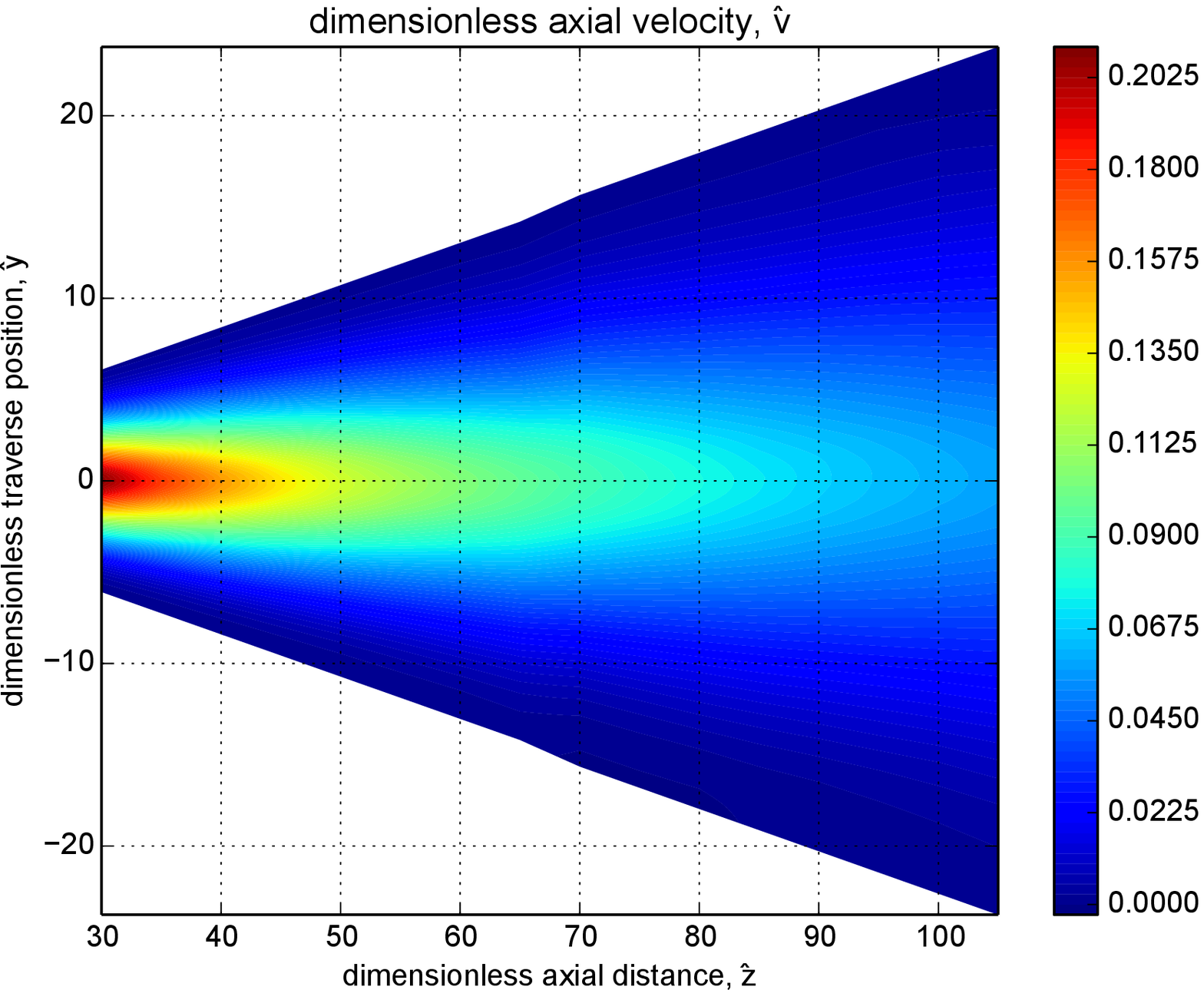}
	\caption{Air jet's axial velocity color map, detailed view of the three different fields. (a) Near field; (b) intermediate field; and (c) far field. The velocity is normalized to the Bernoulli theoretical, Eq. \eqref{eq:v0}, and the distances are normalized by the nozzle's diameter, $D_0$.}
	\label{fig:img-all}
\end{figure*}

\textcolor{black}{Although our main focus is for atomizing liquid jets, the availability of the experimental data for a single-phase air jet made it possible to develop our own original methods for the analysis, as described below, in particular for determining the jet's spread angle, virtual origin and an ``equivalent nozzle diameter''. These original methods  can be analogously applied to the atomizing liquid jet case.} The data for our own experiment at DTU was acquired as described in Sect. \ref{sec:exp}. For these experiments, the initial velocity was $v_{0,exp}=30.59$ m/s, as determined experimentally at the axial position of $\hat{x}=1.16$, remarkably close to the theoretical velocity given by Eq. \eqref{eq:v0}, $v_{0,bern}=30.51$ m/s. The latter value is the one used to normalize the other measured velocities. We estimate $Re_0\approx20,000$. Two data sets were acquired for the single dependent variable of the axial velocity: (i) a coarse 2D field as a function of the axial and traversal positions as coordinates; and (ii) a high-resolution sampling 1D line along the jet's axis. Both of these data sets are divided in the same near, intermediate and far fields. \textcolor{black}{A detailed view of the axial velocity in the different fields is shown in Fig. \ref{fig:img-all}.}

\begin{figure}
	\centering
	\includegraphics[width=\columnwidth]{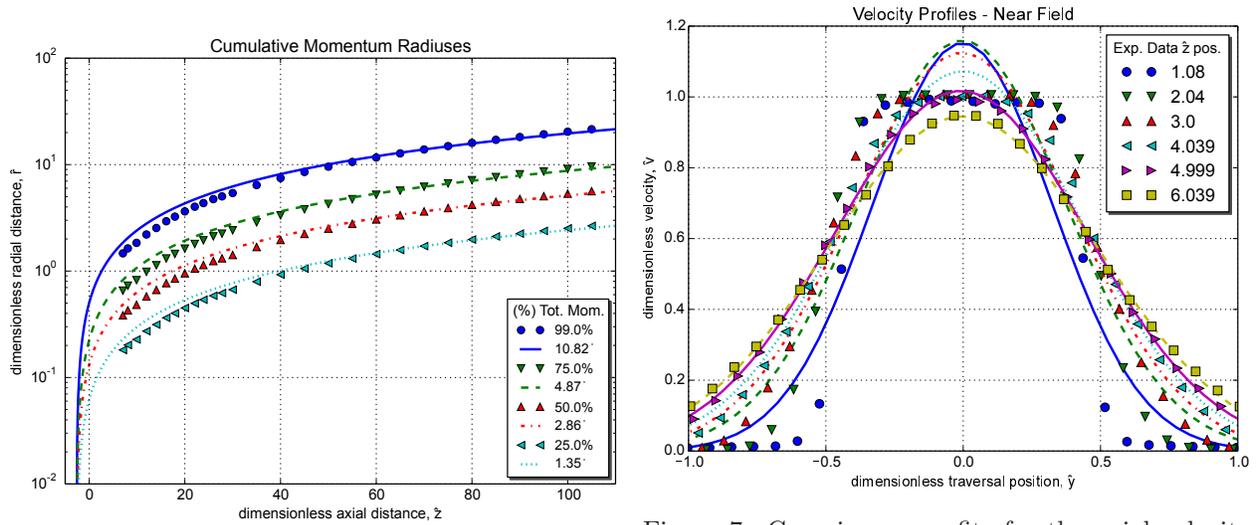}
	\caption{Cumulative Momentum Radii (CMRs) of the air jet and corresponding angle fits (in degrees).}
	\label{fig:CMRs}
\end{figure}

\begin{figure}
	\centering
	\includegraphics[width=\columnwidth]{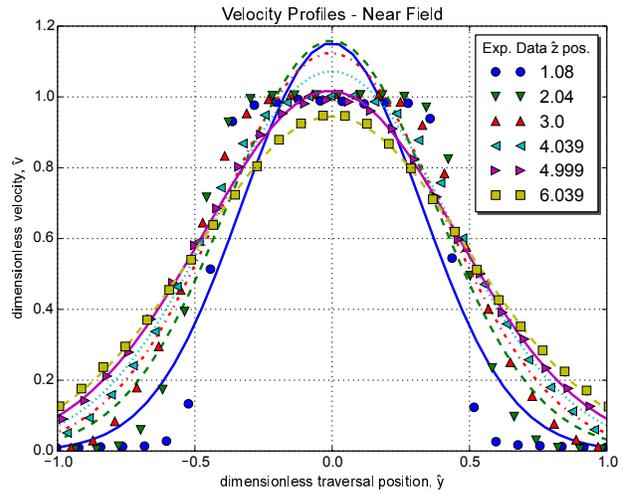}
	\caption{Gaussian curve fits for the axial velocity profiles of the air jet. The velocity is normalized to the Bernoulli initial velocity, $v_0$. Near field, including actual experimental data.}
	\label{fig:GP-near}
\end{figure}

From the Fig. \ref{fig:img-all}(a) we can see there is a small jump or acceleration in the axial velocity of the fluid after exiting the nozzle, reaching a maximum at about 3 nozzle diameters and decaying thereafter. This initial acceleration has been reported by several authors in experiments \citep{Rabadi-etal07, Birouk-Lekic2009} and, in this case, it is most likely due to the expansion of the gas as it comes out of the pressurized conditions inside the nozzle. Our current model does not intend to address this initial acceleration phenomenon, by which we will hereafter only consider the intermediate and far fields. In order to do this, we need to calculate an equivalent nozzle diameter, $d_{eq}$, for a further downstream axial position. This was done by inspecting the jet's geometry and determining an appropriate angle. However, this is not a straightforward task, as there are many possible angle definitions, depending on the focus of different authors \citep{Lefebvre89,Xie2015,Emberson2016}. A common procedure based on the velocity distribution, as done by \cite{Desantes-etal11} is to define the ``half-velocity angle'', i.e. the angle for which the velocity has decayed to half the centerline maximal value. This, however, does not give a full account of the whole limits of the jet, \textcolor{black}{so we here propose} to select an angle that contains 99\% of the total momentum of the jet. In order to do this, we have defined and calculated ``Cumulative Momentum Radii'' (CMRs), i.e. for each axial position, we calculate the radius that contains a certain proportion of the jet's total momentum in that cross section. These CMRs are shown in Fig. \ref{fig:CMRs}. Note that the near field has not been taken into account to calculate the CMRs, the reason being that CMRs were calculated using Gaussian curve fits for each velocity profile at every axial position, and such curves are not a good approximation to the velocity profile in the near field, as can be seen in Fig. \ref{fig:GP-near}. It is worth noting that the exit velocity at the nozzle is very close to that predicted by the Bernoulli theorem ($v_{0,\,bern}=30.51$), as can also be seen in the Fig. \ref{fig:GP-near}, as the velocity used for normalization is precisely $v_0$ which is calculated from the Bernoulli theorem in Eq. \eqref{eq:v0}. From the same figure, we observe that the Gaussian profiles are a reasonable approximation only after about 5 or 6 nozzle diameters; before that, the central portion is evidently flat and the profile is more consistent with a ``top-hat'' shape.
The intermediate field dataset starts at $\hat{x}\approx 7$ (as shown in Fig. \ref{fig:img-all}-b) so the Gaussian profiles are a reasonable approximation thereafter for these datasets, although they tend to underestimate the velocity at the tails. This, however, can be neglected for the purpose of calculating the CMRs as they contain a very small proportion of the total momentum. In the intermediate field, \textcolor{black}{the Gaussian fits are a good approximation overall.} The lines shown in Fig. \ref{fig:CMRs} are least-squares linear fits (shown in logarithmic scale). The CMR99 (99\% CMR) linear fit was calculated by fitting a line with 0.5 as its intercept (one radius), forcing it to cross at the nozzle's exit edge. This defined a virtual origin of the jet inside the nozzle at $\hat{x_0}=-2.62$ (and $\hat{y}=0$); this virtual origin was used as an anchor for the fitting of the lines for subsequent CMRs, so as to produce a consistent set of lines with the same virtual origin defining the concentric cones geometry of the jet.

\begin{figure}
	\centering
	\includegraphics[width=0.7\columnwidth]{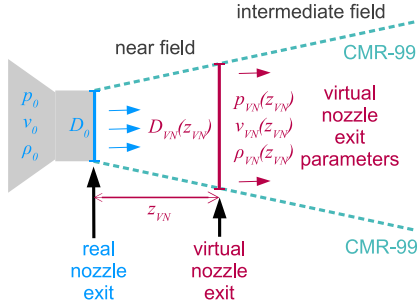}
	\caption{Diagram of the virtual nozzle exit definition and associated dynamical parameters.}
	\label{fig:virtual-nozzle}
\end{figure}

The initial acceleration of the velocity in the near field, as shown in Fig. \ref{fig:img-all}-a, is a physical phenomenon that was not intended to be dealt with in the present ideal momentum jet model, by which we cannot apply the theory to this initial portion of the experimental data. One solution to this is to skip the near field range and define a ``virtual nozzle exit'' at some distance downstream from the real nozzle exit, as shown in Fig. \ref{fig:virtual-nozzle}. We thus define the position of the virtual nozzle exit as the first value of the intermediate field dataset (leftmost value of the horizontal axis in Fig. \ref{fig:img-all}-b), and denote it by $\hat{x}_{VN}=7.06$. With this virtual nozzle exit position, we can define the diameter of the virtual nozzle exit using the CMR99 line shown in Fig. \ref{fig:CMRs}, obtaining $\hat{d}_{VN} = 1.85$. In the following, we take $\hat{d}_{VN}$, the corresponding experimental axial velocity, denoted by $\hat{v}_{VN}=0.88$ (or $v_{VN}=26.88$ m/s), and initial position for the intermediate field to be the boundary values ``at the virtual nozzle exit'' for input in our dynamical models, so as to avoid the initial accelerating region affecting the calculations. We will thus re-normalize our calculations using these values, normalizing distances by $\hat{d}_{VN}$, velocities by $\hat{v}_{VN}$ and shifting axial positions by $\hat{x}_{VN}$.

\begin{figure}
	\centering
	\includegraphics[width=\columnwidth]{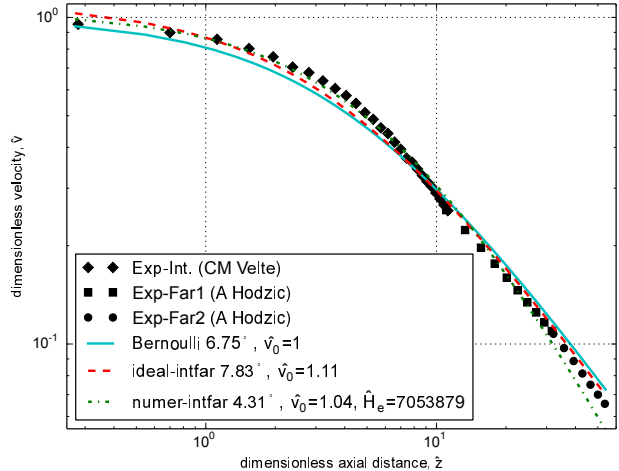}
	\caption{Axial velocity of the jet vs. axial distance. Results for a pressure-atomized single-phase turbulent air jet.}
	\label{fig:jetvel}
\end{figure}

With the above virtual nozzle setting, the evolution of the jet's centerline axial velocity is shown in Fig. \ref{fig:jetvel}. The vertical scale is normalized to $v_{VN}$. All the parameter optimizations were done using only the intermediate and far fields data; in this case the high resolution centerline dataset was used. The ideal momentum submerged jet curve given by Eq. \eqref{eq:mom-jet} is shown as the dashed curve with the label ``ideal-intfar'', which shows excellent agreement with the experimental data. In the same figure, the solid curve shows also Eq. \eqref{eq:mom-jet}, but having fixed the initial velocity to $v_{VN}$ and optimizing the angle only. The other curve shown in a dash-dot line is the result of a model not described in this paper but also developed by the present authors; this \textcolor{black}{``lossy energy jet model''} and other additional related models will be described in detail in a sequel article. Suffice it to say now that an extension of the present model using partial energy conservation by introducing an energy loss factor, which effectively \textcolor{black}{acts} as a simple turbulence model, and producing a numerical scheme (shown as the dash-dot line in Fig. \ref{fig:jetvel}), gives the best result in terms of accuracy, but the small increase in accuracy is at \textcolor{black}{the expense of losing }analytical solutions.

\begin{table}
	\centering
	\caption{Least-squares optimized parameters for the model fittings (except ``*'' which has been fixed before-hand) and the closes corresponding cumulative momentum radius (CMR).}
	\label{tab:optpar}
	\begin{tabular}{l|c|c|c|c}
		Model 			& $\theta\,(^{\circ})$ 	& CMR	& $\hat{v_0}$ 	& $\hat{H_e}$\\
		\hline
		Ideal Momentum 		& 7.83 	& 96	& 1.11 & -\\
		Id.Mo. Bernoulli 	& 6.75 	& 93	& 1.0* & -\\
		\textcolor{black}{Lossy Energy}		& 4.31	& 80	& 1.04 & 7,053,879\\
	\end{tabular}
\end{table}

\begin{figure}
	\centering
	\includegraphics[width=\columnwidth]{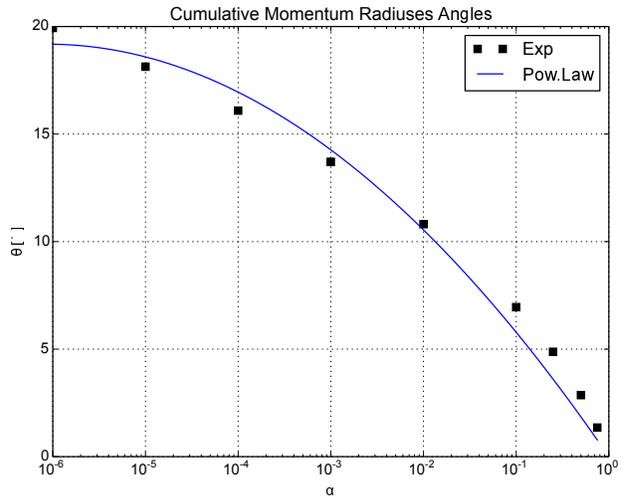}
	\caption{Half-angle of the jet vs. $\alpha$ tail value of the CMR. Results for a pressure-atomized single-phase turbulent air jet.}
	\label{fig:jetangle}
\end{figure}

The results of all parameter optimizations by least-squares are shown in Table \ref{tab:optpar}, where $\hat{H}_e$ is the dimensionless ``energy half-loss'' parameter used in the \textcolor{black}{lossy} energy jet model. Overall, the models give similar results for the initial velocity, where it was left as a free parameter to be optimized, and the angles are between 4 and 8$^{\circ}$. Focusing on the presently discussed model of the ideal momentum jet, when $v_0$ was left as a free parameter, the optimization result was still a velocity close to that of the theoretical Bernoulli velocity (11\% greater). Fixing $v_0$ or setting it as a free parameter gave a similar optimized jet half-angle of around 7$^{\circ}$, which is somewhat lower than the experimental result by \cite{Hosoya-Obokata93}, but still in the range of other experimental results \citep{Lefebvre89,Xie2015,Emberson2016}. In order to determine the best angle for the model optimization, we investigated the relationship between the cumulative momentum value, denoted by $\Pi_c$, and the jet's spread half-angle, $\theta$. Let $\alpha=1-\Pi_c $ so that for the 75\% CMR, $\alpha=0.25$, for the CMR99, \textcolor{black}{$\alpha=0.01$}, etc. The relationship between $\theta$ and $\alpha$ for the turbulent air jet is shown in Fig. \ref{fig:jetangle}, where an experimental curve
\begin{equation}\label{eq:theta}
	\theta(\alpha)=a(1-\alpha^b)+c(1-\alpha^d)
\end{equation}
optimized by a least squares procedure has produced $a\approx-376474$, $b\approx-4.8\times10^{-4}$,  $c\approx112997$, $d\approx-1.6\times10^{-3}$. Based on Eq. \eqref{eq:theta} we can select CMRs closest to a determined half-angle as shown in Table \ref{tab:optpar}. Thus, as a rough approximation we can recommend the CMR95, i.e. the cone angle containing approximately 95\% of the jet's total momentum, as an appropriate angle to be used as a parameter in the current ideal momentum jet model. For this experiment, the CMR95 corresponds to $\theta\approx7.3^\circ$.

\section{Conclusions}\label{sec:conclusions}

We have presented a 1D mathematical model applicable to the dynamics of a wide class of turbulent \textcolor{black}{round} jets. The model's main assumptions include the so-called Locally Homogeneous Flow (LHF) for a two-phase flow. The model is based on conservation laws of the momentum and mass, and describes the dynamical quantities, viz. density, velocity and dynamic pressure, along the jet's axis. The main advantages of the model over others in the literature are that the solutions are analytical, it contains a single free parameter, viz. the jet's angle, and the fact that this angle can be approximated from experimental measurements with an also herein proposed method \textcolor{black}{(Sect. \ref{sub:airjet})}. In particular, we theoretically derive a new explicit formula on a sound physical ground, Eqs. \eqref{eq:me-explicit} and \eqref{eq:Ke-general} for the \textcolor{black}{mass entrainment rate}. Special cases of the model give constant approximations of the \textcolor{black}{mass} entrainment rate coefficients for the near and far fields. The comparison with experimental data for atomizing jets from the literature shows reasonable agreement \textit{in the intermediate and far fields} but more data is needed for testing the limits of the theory. Other special cases considered include a liquid jet in a very thin atmosphere and a submerged jet. In the latter case, the derived state equations agree with the classic result for the velocity decay given by \cite{Landau}. For the same special case, we carried out our own experiments with turbulent air jets, showing excellent agreement with the centerline velocity decay in the intermediate and far fields. A reasonable approach to finding the jet's angle has also been given, introducing the concept of Cumulative Momentum Radii (CMRs).

The present theory can be extended to include partial conservation of energy and mass, producing semi-analytical and numerical solutions, improving the prediction power, as will be described in a subsequent paper.

\section*{Acknowledgements}
FFM thanks the support of the Ministry of Education, Sports, Science and Technology of Japan, the Bank of Mexico's FIDERH program, the Mexican National Council of Science and Technology (CONACYT), Kyushu University's Graduate School of Mathematics and KUMIAY International Co.
This research has also been partially supported by the Center of Innovation Program from the Japan Science and Technology Agency, JST. YF was supported in part by a Grant-in-Aid for Scientific Research from the Japan Society of Promotion of Science (Grant No. 16K05476).
AH and CMV wish to gratefully acknowledge the support of DTU MEK for sponsoring the complete PhD studies program of AH.

\bibliographystyle{plainnat}

\bibliography{IdealJet_Plain2}

\end{document}